\documentclass[conference]{IEEEtran}
\usepackage{amsmath,amssymb,amsbsy,amsxtra,graphicx}
\usepackage{array,multirow}
\usepackage{booktabs}
\usepackage{url}
\usepackage{cite}
\usepackage{bm}
\usepackage{caption}
\usepackage{subfig}
\usepackage{epstopdf}
\usepackage{color}
\usepackage{balance}

\graphicspath{{./Images/}}

\begin{document}

\title{A Compressive Method for Centralized PSD Map Construction with Imperfect Reporting Channel}

\author{\IEEEauthorblockN{Mohammad Eslami, Seyed Hamid Safavi, Farah Torkamani-Azar, Esfandiar Mehrshahi}
\IEEEauthorblockA{Cognitive Communication Research Group, \\Department of Electrical Engineering, \\ Shahid Beheshti University, Tehran, Iran\\
Email: \{m\_eslami, f-torkamani\}@sbu.ac.ir }}

\maketitle

% As a general rule, do not put math, special symbols or citations
% in the abstract
\begin{abstract}
Spectrum resources management of growing demands is a challenging problem and Cognitive Radio (CR) known to be capable of improving the spectrum utilization. Recently, Power Spectral Density (PSD) map is defined to enable the CR to reuse the frequency resources regarding to the area. For this reason, the sensed PSDs are collected by the distributed sensors in the area and fused by a Fusion Center (FC). But, for a given zone, the sensed PSDs by neighbor CR sensors may contain a shared common component for a while. This component can be exploited in the theory of the Distributed Source Coding (DSC) to make the sensors transmission data more compressed. However, uncertain channel fading and random shadowing would lead to varying signal strength at different CRs, even placed close to each other. Hence, existence of some perturbations in the transmission procedure yields to some imperfection in the reporting channel and as a result it degrades the performance remarkably. The main focus of this paper is to be able to reconstruct the PSDs of sensors \textit{robustly} based on the Distributed Compressive Sensing (DCS) when the data transmission is slightly imperfect. Simulation results verify the robustness of the proposed scheme. 

%Spectrum resources are facing huge demands and Cognitive Radio (CR) can improve the spectrum utilization. Recently, Power Spectral Density (PSD) map is defined to enable the CR to reuse the frequency resources regarding to the area. For this reason, the sensed PSDs are fused by a Fusion Center (FC) which the sensed PSDs are collected by the distributed sensors in the area. But, for a given zone, the sensed PSD by neighbor CR sensors may contain a shared common component for a while. This component can be exploited in the theory of the Distributed Source Coding (DSC) to compress sensing data more. In this paper based on the Distributed Compressive Sensing (DCS) a method is proposed to compress and reconstruct the PSDs of the sensors when the data transmission is slightly imperfect. The main focus of this paper is to be able to reconstruct the PSDs of sensors robustly if some perturbations are occurred in the transmission procedure which yields some imperfection in the reporting channel. Simulation results show the advantages of using proposed method in compressing, reducing overhead and also recovering PSDs. 
%% Proposed method can be used to develop a framework when the holding times of the users are large in comparison with the rate of the spectrum sensing.

\end{abstract} 


\IEEEpeerreviewmaketitle

%=======================
\section{Introduction}\label{Introduction}

Nowadays, spectrum resources are facing huge demands by introduction of new emerging communication devices. Cognitive Radio (CR) can tackle with the spectrum scarcity challenge by allowing the secondary users (SUs) to opportunistically access a licensed band while the primary user (PU) is absent. To this end, SUs have to sense the spectrum constantly in order to detect the presence of a primary transmitter signal \cite{Long}. Furthermore, the spectrum sensing has been identified as a key enabling functionality to ensure that cognitive radios would not interfere with PUs, by reliably detecting PU signal. Recently, cooperative spectrum sensing algorithms are proposed to increase the reliability of the spectrum sensing. Correct PU detection probability can be greatly increased by allowing different SUs to share their information and to create collaboration through distributed transmission/processing, in which each user's information is sent out not only by the user, but also by the collaborating users \cite{CRSurvey}. 

Creating an interference map of the operational region at arbitrary locations or frequencies enables the CR to reuse the frequency resources. Hence, it allows a dynamic spectrum allocation scheme \cite{Zhao}. The SUs can keep their transmitted power at the occupied frequency band as low as possible to reuse the frequencies without suffering from or causing harmful interference to the primary system. These reusable zones may be estimated by means of a collaborative scheme whereby receiving the CR Sensors (CRSs) cooperate to estimate the distribution of power in space and frequency as well as localize the positions of the transmitting PUs. In \cite{GIANNAKIS}, sparsity assumption is used in CRs in order to distribute spectrum sensing process and PSD map construction by using Lasso and D-Lasso \cite{LASSO} algorithms. 
%The idea of creating the interference map has been used in \cite{LI} in order to build up deterministic power level map with the aim of routing. In addition, in \cite{SHIH} the geographical-spectral pattern in CR network is reconstructed based on the assumption that the pattern is sparse in a certain transform domain. 

%================= Literature survey
%In \cite{16} a two-step compressed spectrum sensing scheme for efficient wideband sensing is proposed.  In \cite{18}, to collect spatial diversity against wireless fading, multiple CRs cooperate and establish consensus among local spectral estimate through running a decentralized consensus optimization algorithm. In \cite{19}, compressive sampling is performed at local CRs to scan wide spectra, and measurements from multiple CR detectors are fused to collect spatial diversity gain, which improves the detection quality, in particular, under fading channels. Moreover, the authors of \cite{20} and \cite{21} have applied matrix completion and joint sparsity recovery to reduce sensing and transmitting requirements and improve sensing results. Specifically, they equipped CR nodes with a frequency selective filter, so the CR nodes sense linear combinations of multiple channel information and report them to the fusion center, next, using matrix completion and joint sparsity recovery algorithms leads to report the occupied channels. 
%==============================================================

It is known that spectrum sensing needs an accurate and fast decision. Existing papers in the literature mostly focus on the collaborative spectrum sensing performance examination when the fusion center (FC) receives and combines all CR reports \cite{cooperative,ghasemi2005collaborative}. The FC has to deal with many reports from sensors and combine them wisely in order to form a PSD map for arbitrary locations in space and frequency domains. Conventional cooperative sensing can incur significant switching delay and synchronization overhead while wide-band spectrum sensing is of great interest since it can reduce these overheads. Moreover, it is known that the wireless channels are subject to fading and shadowing. When SUs experience multi-path fading or happen to be shadowed, the reports transmitted by CR sensors and users are subject to transmission loss. Besides, when the number of sensors is large, this data transmission which is required for PSD map construction can be challenging. Therefore, a novel method is needed to reduce the amount of data transmitted from the sensors to the FC and also deal with the channel imperfection transmission.

Further benefits such as reducing sensing time and number of sensing measurements with the sub-Nyquist sampling rate can be achieved by employing compressive wide-band spectrum sensing \cite{cooperative}. Compressive Sensing (CS) provides a simultaneous sensing and compression framework \cite{CS}, enabling a potentially significant reduction in the sampling and computation costs at a sensor with limited capabilities. CS technique as the data acquisition approach in a WSN can significantly reduce the energy consumed in the process of sampling and transmission through the network, and also lower the wireless bandwidth required for communication. CS scheme also is exploited in PSD map construction \cite{MyIEICE} as well as many other research fields.

The DCS concept was merely used for wideband spectrum sensing in a reliable state of the art works. A distributed compressive wide-band spectrum sensing is proposed in \cite{CrJSM1} and shows that the performance gains arising from the use of spatial diversity as well as joint sparsity.  
%In \cite{CrJSM4}, the authors develop a multirate asynchronous sub-Nyquist sampling system that employs multiple low-rate analog-to-digital converters (ADCs). 
%Also, \cite{CrJSM5} proposes two solution algorithms for its simple DCS based spectrum sensing. 
A novel Analog-to-Information Converters (AIC) structure of CR front-end integrating low rate Analog-to-Digitial Converters (ADC) and few storage units is proposed in \cite{CrJSM2} while authors also explore a new joint sparsity model in CR networks and provide a solution algorithm to perform joint spectrum reconstruction. 
To reduce the transmitted data, our proposed framework in \cite{Mine} is different from the state-of-the-art papers in \cite {CrJSM1,CrJSM5,CrJSM2,CrJSM4} in some aspects. They suppose an special form of DCS (JSM-2) and developed their algorithms with respect to it. In addition some of them needs some other information such as the sparsity level of the spectrum too. While the JSM-2 is not necessarily the best form to deal and exploit in wideband spectrum sensing. The proposed framework in \cite{Mine} is independent form these assumptions.

It is important to notice that while joint spectrum recovery requires much less samples for each CR, uncertain channel fading and random shadowing would lead to varying signal strength at different CRs, even placed close to each other. Different from \cite{Mine}, the main focus of this paper is to be able to reconstruct the PSDs of sensors robustly if some perturbations are occurred in the transmission procedure which yields some imperfection in the reporting channel. Therefore, we model the perturbation of the transmission system by using the disturbance filters between each sensor node and the FC. The reconstruction formula of the proposed method includes the estimated version of these filters and it brings some compensations against the occurred errors. In addition, by assuming that the common component is known with both FC and sensors, we propose an scheme to estimate the disturbance filters. We will show that we can greatly improve the performance by considering the reporting channel imperfection \footnote{The two page summary of this work has accepted in PhD forum of ICASSP 2017 to present \cite{Eslami2017ICASSP}. This is to certify that PhD forum papers will not be indexed in IEEE.}.

The rest of the paper is organized as follows: Section \ref{Prob_Form} describes the system model of the cognitive network. Section \ref{Prop_approach} presents the proposed method. Simulation results and discussions are given in section \ref{Experiments}. Finally, section \ref{Conclusion} concludes the paper and some future directions are presented.

%--------------------------------<< Proposed Approach >>-----------------------------------------%
\section{Problem Formulation}\label{Prob_Form}
In compressive spectrum sensing \cite{CrCs3}, the sensed PSD can be performed by 
\begin{equation*}
\boldsymbol s = \boldsymbol W \boldsymbol F \boldsymbol a
\end{equation*}
%==============
where an smoothing operator $\boldsymbol W$ followed by a Fourier transform $\boldsymbol F$ are  exploited on the auto-correlation vector $\boldsymbol a$ of the sensed signals. Since using smoothing operator, the PSD is almost piecewise constant generally. Therefore the number of significant non-zero values in the edge vector of PSD $\boldsymbol z$ is so sparse. The edge vector can be found by using $\boldsymbol z = \Gamma  \boldsymbol s$ where in the most simple case $ \Gamma$ is a differential operator as expressed in \cite{CrCs3}. The dictionary of the sparse representation ($\boldsymbol D$) can be found as 
\begin{equation*}
\boldsymbol z = \Gamma  \boldsymbol W \boldsymbol F \boldsymbol a = \boldsymbol D^{-1} \boldsymbol a
\end{equation*}
Now, reconstructing the sparse edge vector of the PSD is possible from the sensed measurements $\boldsymbol y = \Phi  \boldsymbol a$. Finally, the estimated PSD, $\hat{\boldsymbol s}$, can be achieved by 
\begin{equation*}
\hat{\boldsymbol s} = \boldsymbol G \hat{\boldsymbol z}
\end{equation*}
where $\hat{\boldsymbol z}$ is the estimated edge vector and $\boldsymbol G $ is the cumulative sum matrix (i.e. a lower triangular matrix with $+1$ elements). 

Suppose that $M^2$ sensors should capture a PSD $\boldsymbol s_j \in R^N$ ($j \in {1,2,..., M^2}$) and are distributed in the area. It can be expected that there is a shared common component $\boldsymbol s_c \in R^N$ between the $J$  neighbor sensors which constitute a Group of Sensors (GoS), such that 
\begin{equation*}
\boldsymbol s_j=\boldsymbol s_c + \boldsymbol s_{{inn}_j}
\end{equation*}
where $\boldsymbol s_{{inn}_j} \in R^N$ is the innovation part of each PSD $\boldsymbol s_j$. PSDs can be represented as a cumulative sum of their edge vector as 
\begin{equation*}
\boldsymbol s_j= \boldsymbol G \boldsymbol z_{j}
\end{equation*}
by matrix $\boldsymbol G \in R^{N\times N}$. Obviously, we have sparse $\boldsymbol z_c$ and $\boldsymbol z_{{inn}_j}$s which belong to space $R^N$ with different sparsity levels. Therefore, 
\begin{equation*}
\boldsymbol s_j=\boldsymbol G \boldsymbol z_{j}=\boldsymbol G (\boldsymbol z_c+\boldsymbol z_{{inn}_j})
\end{equation*}
and
\begin{equation*}
\boldsymbol s_c=\boldsymbol G \boldsymbol z_c, \boldsymbol s_{{inn}_j}=\boldsymbol G \boldsymbol z_{{inn}_j}
\end{equation*}
$\Phi_j \in R^{w_j \times N}$ is an individual measurement matrix for the $j$th sensor and its sensed measurements $\boldsymbol y_j = \Phi_j \boldsymbol a_j$ should be sent to the FC. $\boldsymbol r_j \in R^{w_j}$ is the received signal by the FC according to the sent $\boldsymbol y_j$. Notice that in the rest of the paper the $\hat{}$ denotes the reconstructed vectors or signals. 

In this document, we seek to propose criteria for compressing the sensed PSDs of the sensors to reduce the amount of the transmitted data from sensors to FC and also deal with the channel imperfection transmission.

%In order to summarize and also analyze the overhead and gain of the following proposed method, a simple cognitive timing structure in the worst case is assumed in which the sensors are the cognitive users too, transmissions should be passed by FC and also there is no control channel. The details of the assumed framework is shown Fig. \ref{overhead} with respect to the time structure of each sensor. The main message of each cognitive user will be sent in the \textit{Periodic usage phase} ($P$ time slots with duration $T$) after that the initial permission of using and also how using is decided by FC and fedback to the sensors in \textit{Initialization phase}$Ti$. Sensors sense the channels ( $t_s$) and report their measurements to FC ($t_r$). Then, sensors wait for a decisions by FC ($t_w$) and listen to the FC's commands ($t_l$) in \textit{Initialization phase}. The main message (data) of the sensors (CR users) send in $t_t$ in \textit{Periodic usage phase}. The efficiency in each time slot can be interpreted as $t_t/T$. Proposed method brings lower sensing rate $w_j$ and so makes $t_{r'} < t_r, \ t_{s'} < t_s$ and therefore $t_{t'} > t_t$ which brings more gain. 

%\begin{figure}
%  \centering
%  \includegraphics[width= .4 \textwidth]{FigOverhead.png}
% \caption{A simple time structure for cognitive usage. up: initialization phase, down: periodic spectrum usage.}
%  \label{overhead}
%\end{figure}

%--------------------------------<< Proposed Approach >>-----------------------------------------%
\section{Proposed Approach}\label{Prop_approach}
\subsection{Perfect Reporting Channel ($\boldsymbol{r}_j = \boldsymbol{y}_j$)}
The simplest idea to reconstruct the PSD $\boldsymbol s_j$ is $\hat{\boldsymbol s}_j=  \boldsymbol G \hat{\boldsymbol z}_j$ where $\hat{\boldsymbol z_j}$ is attained by solving the BPDN \cite{BPDN} inspired problem eq. \eqref{eq.00}. Also we can use the shared common component $\boldsymbol s_c$ in a GoS based on the Joint Sparsity model (JSM) \cite{DCS} and therefore, reconstruct the data in lower measuring (sensing) rate. Inspired from JSM, equations (\ref{eq.10}) to (\ref{eq.13}) are defined to model recontsruction for a $J$ neighbor sensors (a GoS). Therefore, the desired PSDs can be yielded by 
\begin{equation}
\hat{\boldsymbol s}_j=\boldsymbol G (\hat {\boldsymbol z}_c + \hat{{\boldsymbol z}}_{{inn}_j})
\end{equation}
%====================
where $\hat {\boldsymbol z}_c$ and $\hat{{\boldsymbol z}}_{{inn}_j}$s are located in the found $\hat{\boldsymbol z}  = {\left[ {\begin{array}{*{20}{c}} {{\hat{\boldsymbol z} _c}^T}&{{{\hat{\boldsymbol z}}_{{inn}_1}}^T}& \cdots &{{{\hat{\boldsymbol z}}_{{inn}_J}}^T} \end{array}} \right]^T}$ vector. $\hat{\boldsymbol z}$ is computed by solving the optimization problem in equation (\ref{eq.14}) where $\boldsymbol r \in R^{W}$, $\boldsymbol n \in R^{W}$, $\boldsymbol z \in R^{N(J+1)}$, $\Psi \in R^{W \times N(J+1)}$ and $W = \sum\limits_{j = 1}^J {{w_j}}$.
\begin{align}
\hat{\boldsymbol z}_j= \mathop {\min }\limits_{\acute{\boldsymbol z}_j}   \frac{1}{2} \left\| {\boldsymbol r_j- \Phi_j D \acute{\boldsymbol z}_j\|} \right._2^2 +  \lambda \left\|\acute{\boldsymbol z}_j \right\|_1. \label{eq.00} \ \ \ \ \ \ \ \\
\boldsymbol r= \Psi \boldsymbol z + \boldsymbol n \label{eq.10} \ \ \ \ \ \ \ \ \ \  \ \ \ \ \ \ \ \ \ \ \ \ \\
\boldsymbol r = \left[ {\begin{array}{*{20}{c}}  
{{\boldsymbol r_1}^T} \ 
{{\boldsymbol r_2}^T} \ 
 \hdots \ 
{{\boldsymbol r_J}^T}
\end{array}} \right]^T \ \ \ \ \ \ \ \ \ \ \ \nonumber \\ \boldsymbol n = \left[ {\begin{array}{*{20}{c}}
{{\boldsymbol n_1}^T} \ 
{{\boldsymbol n_2}^T} \ 
 \hdots \ 
{{\boldsymbol n_J}^T}
\end{array}} \right]^T \ \ \ \ \ \ \ \ \ \ \label{eq.11} \\  \boldsymbol z = \left[ {\begin{array}{*{20}{c}}
{{\boldsymbol z_c}^T} \ 
{{\boldsymbol z_{{inn}_1}}^T} \ 
 \hdots \ 
{{\boldsymbol z_{{inn}_J}}^T} \ 
\end{array}} \right]^T \ \ \ \ \ \ \ \nonumber \\
\Psi  = \left[ {\begin{array}{*{20}{c}}
{\Phi _1} \boldsymbol D& {\Phi _1} \boldsymbol D&{\begin{array}{*{20}{c}}
0& \cdots 
\end{array}}&0\\
{\Phi _2} \boldsymbol D&0&{\begin{array}{*{20}{c}}
{\Phi _2} \boldsymbol D& \cdots 
\end{array}}& \vdots \\
 \vdots &{}&{\begin{array}{*{20}{c}}
{}& \ddots 
\end{array}}&0\\
{\Phi _J}\boldsymbol D&0&{\begin{array}{*{20}{c}}
 \cdots 
\end{array}}&{\Phi _J}\boldsymbol D
\end{array}} \right] \label{eq.13} \\
\hat{\boldsymbol z}= \mathop {\min }\limits_{\acute{\boldsymbol z}}   \frac{1}{2} \left\| {\boldsymbol r - \Psi \acute{\boldsymbol z}\|} \right._2^2 +  \lambda \left\|\acute{\boldsymbol z}\right\|_1  \ \ \ \ \ \ \ \ \label{eq.14}
\end{align}

But, assume a scenario in which the common part $\boldsymbol z_c$ is known by the FC and fix for a while of time. For example, the holding time is 30 seconds while the network is sensed in each 0.3 seconds. Here, in order to enhance the model, we remove the common part from the reconstruction equation. Equation (\ref{eq.10}) can be rewritten in form of equation (\ref{eq.16}) and split by using a combination of two distinct parts: Common part $\boldsymbol z_c$ and Innovation part $\boldsymbol z_I$. The reconstruction formula can be modified to just find the innovation parts of the PSDs by \eqref{eq.17} where $\boldsymbol r_{inn}=\boldsymbol r-\boldsymbol A \boldsymbol z_c$. Consequently, the PSD of each sensor will be found by 
\begin{equation}
\hat{\boldsymbol s}_j= \boldsymbol G (\boldsymbol z_c + \hat{{\boldsymbol z}}_{{inn}_j})
\end{equation}
%===============
where $\hat{{\boldsymbol z}}_{{inn}_j}$s are located in the computed $\hat{{\boldsymbol z}}_I$ vector. This modification brings faster solution and also better reconstruction accuracy. 
\begin{align}
\boldsymbol r=\left[ {\boldsymbol A\| \boldsymbol H} \right] \left[ {\begin{array}{*{20}{c}}
{{\boldsymbol z_c}^T} \ 
{{\|}} \ 
{{\boldsymbol z_{{inn}_1}}^T} \
 \hdots \ 
{{\boldsymbol z_{{inn}_J}}^T}
\end{array}} \right]^T + \boldsymbol n \nonumber \\ = \boldsymbol A \boldsymbol z_c +\boldsymbol H \boldsymbol z_I +\boldsymbol n  \label{eq.16} \ \ \ \ \ \ \ \ \ \ \ \ \\ 
\hat{{\boldsymbol z}}_I= \mathop {\min }\limits_{\acute{\boldsymbol z_I}}   \frac{1}{2} \left\| {\boldsymbol r_{inn} - \boldsymbol H \acute{\boldsymbol z_I}\|} \right._2^2 +  \lambda \left\|\acute{\boldsymbol z_I}\right\|_1 \label{eq.17} \ \ \ \
\end{align} 

Now we try to find the optimum common component $\boldsymbol z_c$ labeled as $\boldsymbol z_{c_{opt}}$. Since our proposed model is based on JSM, the first well known approach to find the optimum is solving the JSM based optimization problem \eqref{eq.19} where $\overline {\boldsymbol G}$ is a matrix constructed by arranging $\boldsymbol G$s similar to eq. \eqref{eq.13} ($\phi_j$s are replaced by $G$s). But remember that the desired variables in the proposed eq. (\ref{eq.17}) are just innovation parts $\boldsymbol z_I$ and the optimization constraint is just the maximum sparsity of the innovation parts. Therefore the problem to find the $\boldsymbol z_{c_{opt}}$ can be exchanged to \eqref{eq.20} where the $\boldsymbol z_{c_{opt}}$ is a part of the found  $\boldsymbol z_{opt}$ vector. The more details and achievements of these proposed criteria are published in \cite{Mine}.
\begin{align}
\label{eq.18}
\boldsymbol z_{opt} = \left[ {\begin{array}{*{20}{c}}
{{\boldsymbol z_{c_{opt}}}^T} \ 
{{\boldsymbol z_{{inn}_{1_{opt}}}}^T} \ 
 \hdots \ 
{{\boldsymbol z_{{inn}_{J_{opt}}}}^T}
\end{array}} \right]^T \nonumber \\ \boldsymbol s_{all} = \left[ {\begin{array}{*{20}{c}}
{{\boldsymbol s_1}^T} \ 
{{\boldsymbol s_2}^T} \ 
 \hdots \
{{\boldsymbol s_J}^T} 
\end{array}} \right]^T \ \ \ \ \ \ \ \ \ \  \\
\boldsymbol z_{opt}= \mathop {\min }\limits_{\acute{\boldsymbol z}_{opt}} \left\|\acute{\boldsymbol z}_{opt}\right\|_{1} subject \ to \  \boldsymbol s_{all} =  \overline {\boldsymbol G} \acute{\boldsymbol z}_{opt}  \label{eq.19} \\
%\end{align}   
%\begin{align}
\boldsymbol z_{opt}= \mathop {\min }\limits_{\acute{\boldsymbol z}_{opt}} \sum\limits_{j = 1}^J \left|{\acute{\boldsymbol z}_{{inn}_{j_{opt}}}}\right|_1 subject \ to \  \boldsymbol s_{all} =  \overline {\boldsymbol G}  \acute{\boldsymbol z}_{opt} \label{eq.20}
\end{align}
\subsection{Imperfect Reporting Channel ($\boldsymbol{r}_j \ne \boldsymbol{y}_j$)}
Now let us use the known and fixed for a while common part $\boldsymbol z_{opt}$ more. The above mentioned criteria was based on our model in \eqref{eq.10}. But, it is clear that, recovering the PSDs by using eq. (\ref{eq.00}) and also eq. (\ref{eq.14}) only can be useful when there is no signifant difference between $\boldsymbol r_j$ and $\boldsymbol y_j$. In order to embed the effect of this imperfection in the model, we have proposed to estimate a destructive filter $\boldsymbol \beta_j \in R^{w_j}$ and exploit it with the model by circular convolution as 
\begin{equation*}
\boldsymbol r_j=\boldsymbol y_j \odot \boldsymbol\beta_j+\boldsymbol n_j.
\end{equation*}
Similarly we can model the received signal of the common part $\boldsymbol r_{c_j}$ as 
\begin{equation*}
\boldsymbol r_{c_j}=\boldsymbol y_{c_j} \odot \boldsymbol\beta_j +\boldsymbol \breve{n}_j
\end{equation*}
or in the matrix multiplication form as 
\begin{equation*}
\boldsymbol r_{c_j}= {\mathop {\boldsymbol Y}\limits^o}_{c_j} \boldsymbol\beta_j + \boldsymbol \breve{n}_j
\end{equation*}
where ${\mathop {\boldsymbol Y}\limits^o}_{c_j} \in R^{w_j \times w_j}$ is the circulant matrix of $\boldsymbol y_{c_j} \in R^{w_j}$. More clearly, if 
$\boldsymbol y_{c_j}$ contains samples such that $\boldsymbol y_{c_j} = {[{y_{c_j}}(1),{y_{c_j}}(2), \cdots ,{y_{c_j}}({w_j})]^T}$ then the $\bar{\bar{Y}}_{c_j}$ will be 

\begin{align}
\label{eq.filter3}
\bar{\bar{\boldsymbol Y}}_{c_j}  = \left[ {\begin{array}{*{20}{c}}
{{y_{c_j}}(1)}&{{y_{c_j}}({w_j})}& \cdots &{{y_{c_j}}(2)}\\
{{y_{c_j}}(2)} &{{y_{c_j}}(1)} & \ddots  & \vdots \\
\vdots & \ddots & \ddots  &{{y_{c_j}}({w_j})}\\
{{y_{c_j}}({w_j})}& \cdots&{{y_{c_j}}(2)} &{{y_{c_j}}(1)}
\end{array}} \right]
\end{align}

So if assume that $\boldsymbol y_{c_j}$ signals are known by the FC or equivalently FC knows the $\boldsymbol z_c$ and $\Phi_j$s. Therefore, estimated impulse response of the destructive filter can be achieved by solving the optimization problem 
\begin{equation}
\hat{\boldsymbol\beta}_j= \mathop {\min }\limits_{\acute{\boldsymbol\beta_j}} \left\| \boldsymbol r_{c_j}-{\mathop {\boldsymbol Y}\limits^o}_{c_j} \acute{\boldsymbol\beta_j} \right\|_2
\end{equation}
%===========================
Now, after estimating the destructive filters $\hat{\boldsymbol\beta_j}$s, which are related to the communication paths between the $j$th sensor and the fusion center, we can construct the circulant matrix of the $\hat{\boldsymbol\beta_j}$ as 
\begin{align}
\label{eq.0}
{\mathop {\boldsymbol B}\limits^o}_{j} = \left[ {\begin{array}{*{20}{c}}
{{\hat{\beta}_j}(1)}&{{\hat{\beta}_j}({w_j})}& \cdots &{{\hat{\beta}_j}(2)}\\
 {{\hat{\beta}_j}(2)} &{{\hat{\beta}_j}(1)} & \ddots  & \vdots \\
\vdots & \ddots & \ddots  &{{\hat{\beta}_j}({w_j})}\\
{{\hat{\beta}_j}({w_j})}& \cdots&{{\hat{\beta}_j}(2)} &{{\hat{\beta}_j}(1)}
\end{array}} \right] \in R^{w_j\times w_j}
\end{align}
%===========================
%$\bar{\bar{\boldsymbol B}}_{j} \in R^{(w_j)\times w_j}$ in equation (\ref{eq.0}) and reformulate the eq. (\ref{eq.10}) to equation (\ref{eq.14.7}) where $\overline{\overline {\boldsymbol B}} \in R^{W \times W}$ is defined in (\ref{eq.14.8}).
Reformulating eq. (\ref{eq.10}) brings new model as follows:
\begin{equation}
\boldsymbol r=\overline{\overline {\boldsymbol B}}  \Phi \Psi \boldsymbol z + \boldsymbol n \label{eq.14.7}
\end{equation}
where $\overline{\overline {\boldsymbol B}} \in R^{W \times W}$ is defined as follows:
%and ${\mathop {\boldsymbol B}\limits^o}_{j} \in R^{(w_j)\times w_j}$ is the circulant matrix of the $\hat{\boldsymbol\beta_j}$.
%===========================
\begin{align}
\overline{\overline {\boldsymbol B}}  = \left[ {\begin{array}{*{20}{c}}
{{\mathop {\boldsymbol B}\limits^o}_1 }&0& \cdots &0\\
0&{{\mathop {\boldsymbol B}\limits^o}_2 }& \cdots &0\\
 \vdots & \cdots & \ddots & \vdots \\
0& \cdots &0&{{\mathop {\boldsymbol B}\limits^o}_J}
\end{array}} \right] \label{eq.14.8}
\end{align}
%===========================
The reconstruction criterion for this new model is similar to eq. \eqref{eq.17} while the entities of matrix $\boldsymbol H$ are modified.

%But the proposed method has larger initialization phase duration (waiting overhead) because of more complexity and solution times ($Ti'>Ti$). Obviously, in order to have more gain and lower overhead we should have $Pt_{r'}+Pt_{s'}+t_o<<Pt_{r}+Pt_{s}$ or equivalently have larger $P$. The holding time of the users in the area can be useful to deal with this issue and able us to use our compression efficiency more practically. It can be assumed that the shortest time length in which the structures of the sensed PSDs of the sensors remain unchanged is the holding time ($t_h$) of the channels. Therefore, the following problem can be formulated to show the relations between holding time and our potential gains:
% $argmax \ P, \ T \ subject \ to \ PT<t_h \ , \ P \in \mathbb{N}$. In a typical manner when the holding time is $t_h=20$ seconds and the network is sensed in each $T=0.3$ seconds, we can have approximately $P\approx60$ and this means that the proposed method can bring significant gains somewhere.

%=======================
\section{Experiments}\label{Experiments}
%=======================
In order to simulate the PSD map, the following constraints are considered:
\begin{itemize}
\item The zone consists of $122 \times 122$ grids in which each PU or SU can exist.
\item The zone consists of $M  \times M (M=12)$ uniformly distributed sensors.
\item The PSD of each sensor consists of $32$ frequency subbands while each channel interval contains $8$ samples (therefore, $N=256$).
\item For each frequency subband, at most $50$ PUs are active and located randomly permuted on the grids of the zone.
\item The power of PUs signals vanishes with respect to the distance.
\item We have selected the number of neighbor sensors in each GoS as $J=4$.
\item FC is located in the center of the zone.
\end{itemize}
%===============
%-------------------------------------------------
\begin{figure}[t]
 \centering
 \includegraphics[width=0.5 \textwidth]{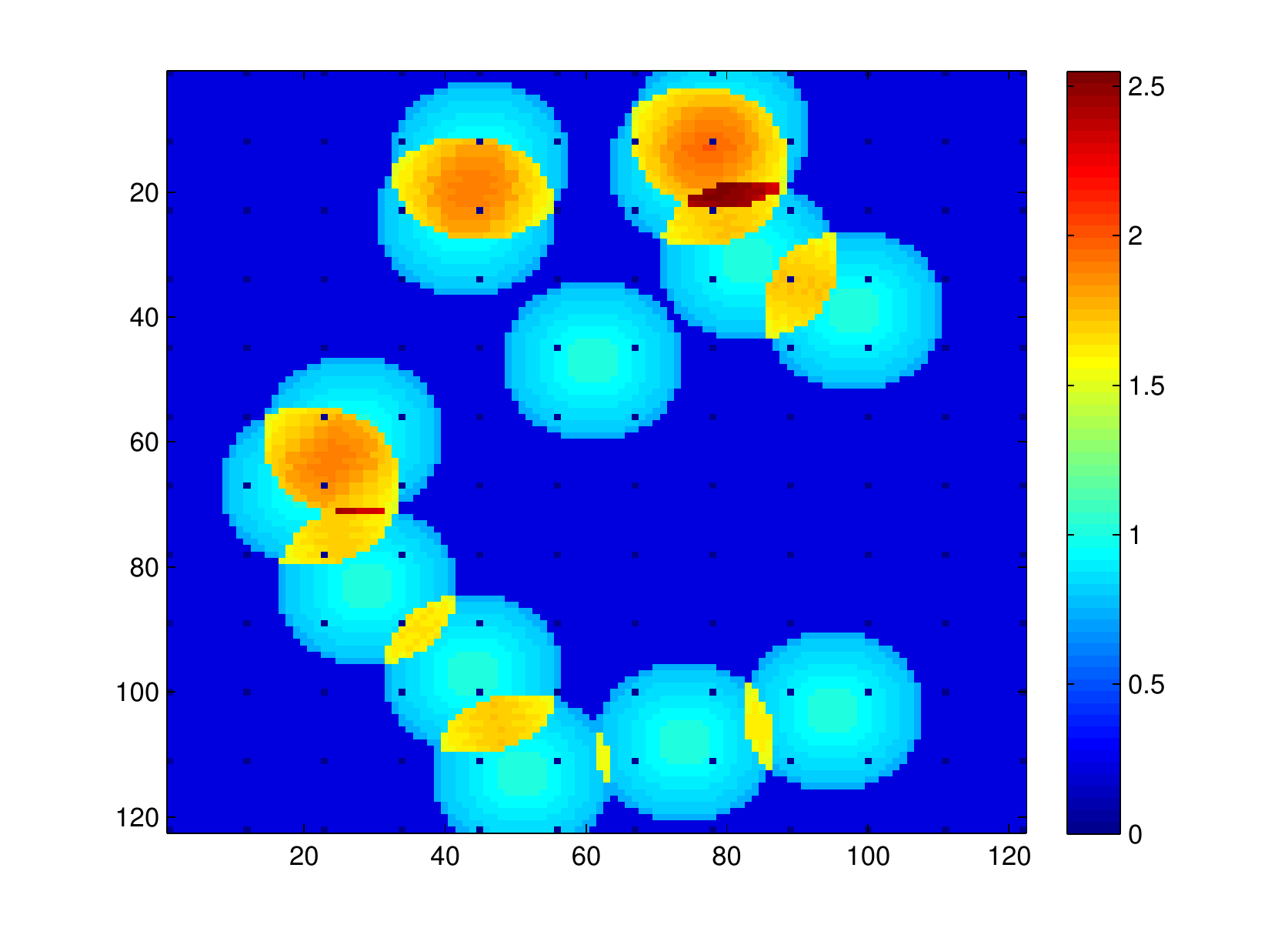}
 \caption{A representation of a PSD Map in a given frequency subband. $122 \times 122$ grid points with $12 \times 12$ uniformly distributed sensors.}
  \label{PSDmap} 
\end{figure}

\begin{figure}
 \centering
 \includegraphics[width=0.5 \textwidth]{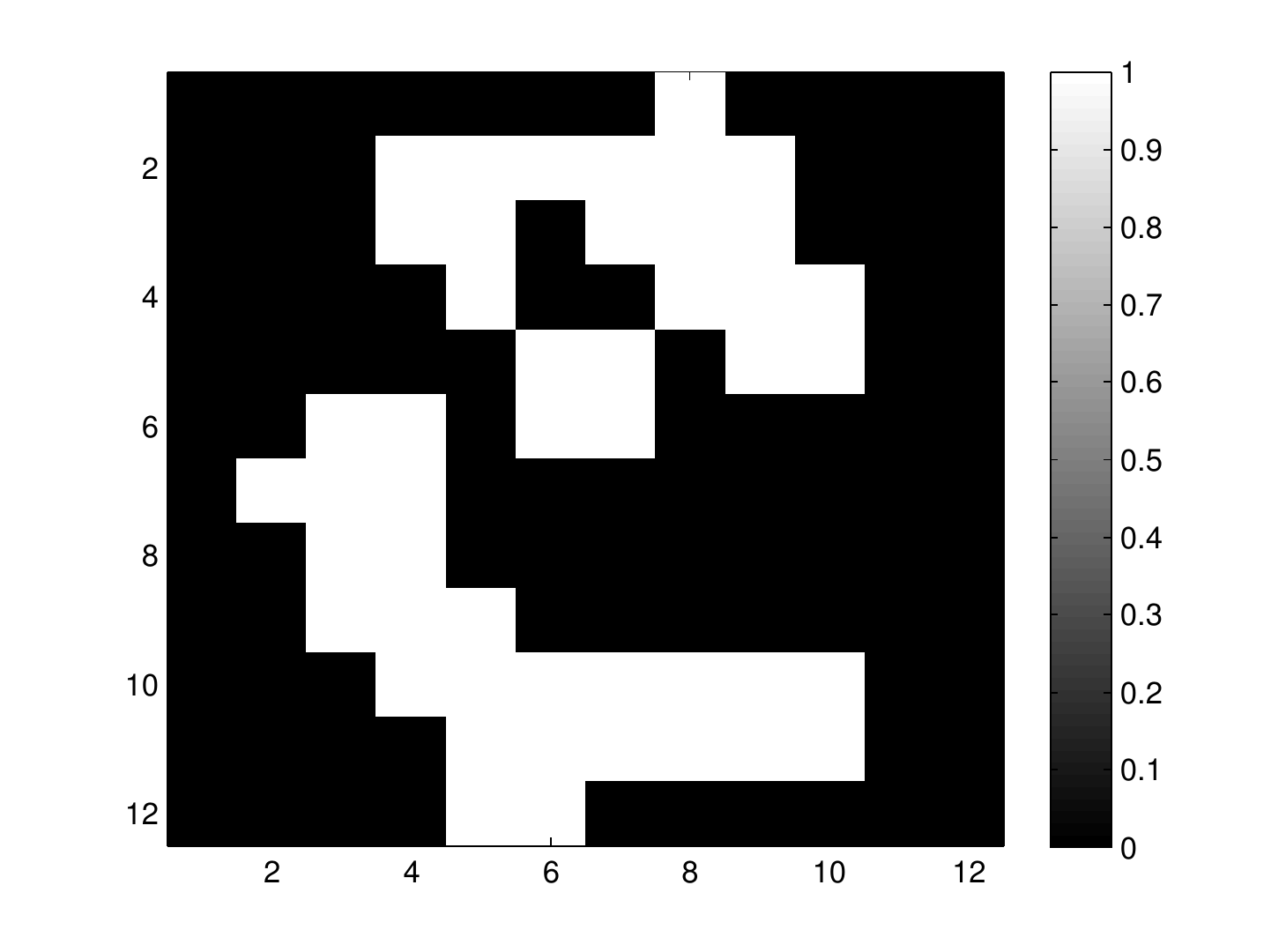}
 \caption{The representation of the occupancy map regarding to the PSD Map of Fig. \ref{PSDmap} for its $12 \times 12$ uniformly distributed sensors}
  \label{PSDmapBinary} 
\end{figure}

By using this mentioned scenario, each sensor captures its individual PSD and therefore the occupancy and un-occupancy of each subband is modeled randomly. Sensors use different measurement matrices $\Phi_j \in R^{w_j \times N}$ with Gaussian random set of projections. The sensed samples $\boldsymbol y_j \in R^{w_j}, j=\{1,2,...,M^2\}$ are sent to the FC through a digital transceiver system. BPSK modulation, $1/2$ channel encoding, and DS-CDMA with $4$-chip's length are the specifications of the used transceiver system. Simulations are experimented for $100$ snapshots of time with random behavior PSD maps and the achieved mean results are reported. We use the SparseLab \cite{BPDN2} Matlab toolbox to run the BPDN algorithm. 

\begin{figure}[b]
  \centering
  \includegraphics[width=9cm, height=6cm]{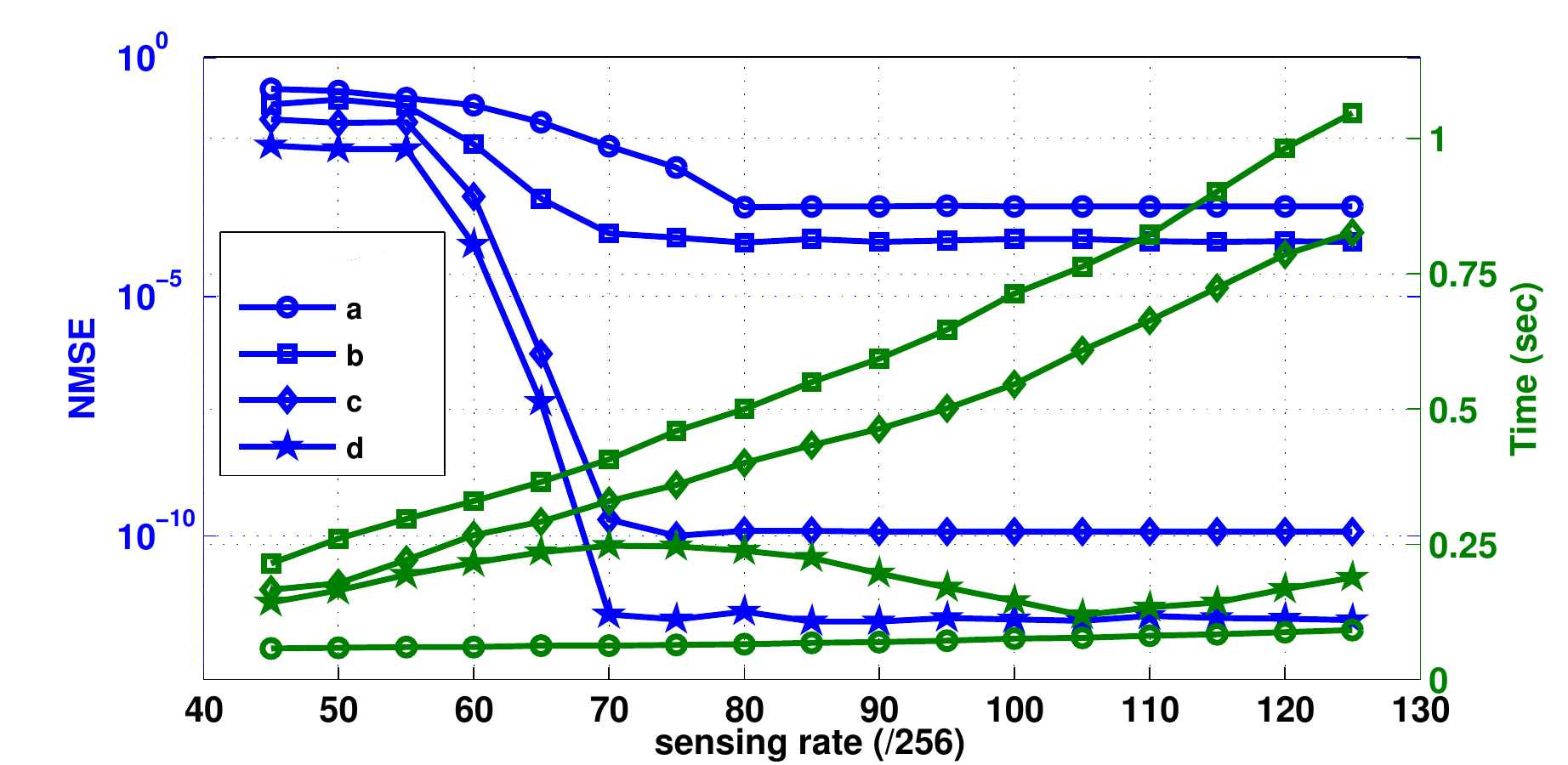}
  \caption{Comparison between compression ability and reconstruction time of the mentioned models. a) Individual reconstruction, eq. (1). b) JSM reconstruction, eq. (5). c) Proposed reconstruction eq. (7) with $\boldsymbol z_{c_{opt}}$ found by eq. (9). d) Proposed reconstruction eq. (7) with $\boldsymbol z_{c_{opt}}$ found by eq. (10) \cite{Mine}}.
  \label{MseIEEE}
\end{figure}

\begin{figure*}
 \centering
 \includegraphics[width=0.8 \textwidth]{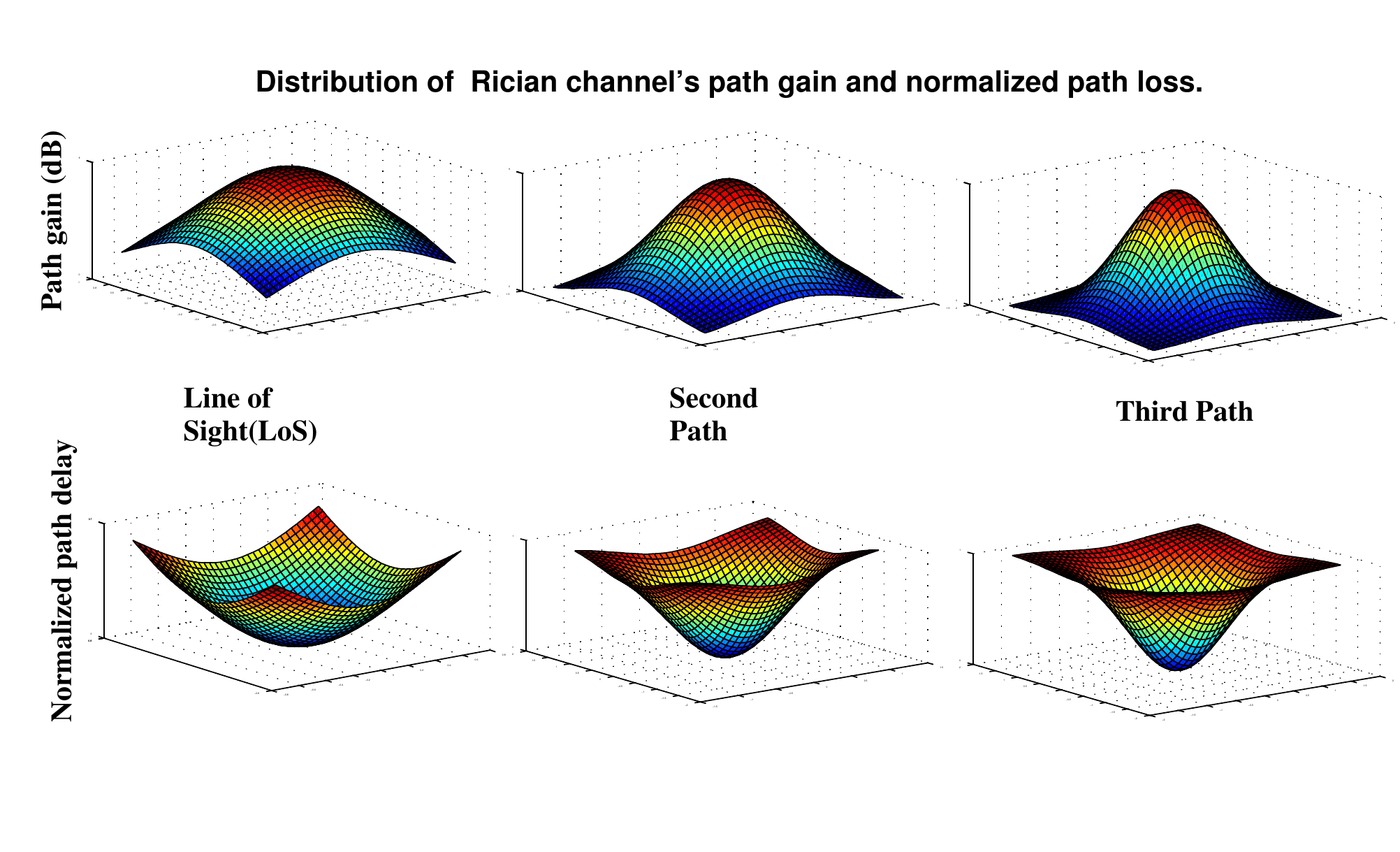}
 \caption{ Spatial distribution of Rician channel's path gains (dB) and normalized path delay: Line of Sight, second and third path, respectively.}
  \label{Pathloss} 
\end{figure*}

By using above mentioned scenario, each sensor capture its individual PSD and therefore the occupancy and un-occupancy of each subband is modeled randomly. Figures \ref{PSDmap}  and \ref{PSDmapBinary} show a PSD map of a subband and its corresponding occupancy map which should be sensed and estimated by the sensors and decided by the FC.
Fig. \ref{MseIEEE} shows the achievements of the proposed method with respect to reducing sensing rate and reconstruction time in a loss-less network. 

In order to consider the effect of imperfect channel, another experimental results are simulated by using a Rician channel with 3 paths by different path-gains and delays depending on the relative locations of the sensors with respect to the FC. The spatial distribution of Rician channel's path gains (dB) and the normalized Rician channel's path delay for each CR sensor are depicted in Fig. \ref{Pathloss}, respectively. The maximum of each path-gains are $\left[10 \times 10^{-1}, \ \ 9.0 \times 10^{-1}, \ \ 8.0 \times 10^{-1}\right]$(dB), the minimum of each path delays are $\left[10^{-015}, \ \ 10^{-017}, \ \ 10^{-019} \right]$ seconds. It is clearly shown that when one CR sensor is far away from the FC, the received signal is very week, therefore the farthest CR sensors' data over the channel experiences the maximum delay and minimum path-gain and vice versa. In addition, we assume an additive white Gaussian noise with different noise levels.

\begin{figure}[t]
 \centering
 \includegraphics[width=0.5 \textwidth]{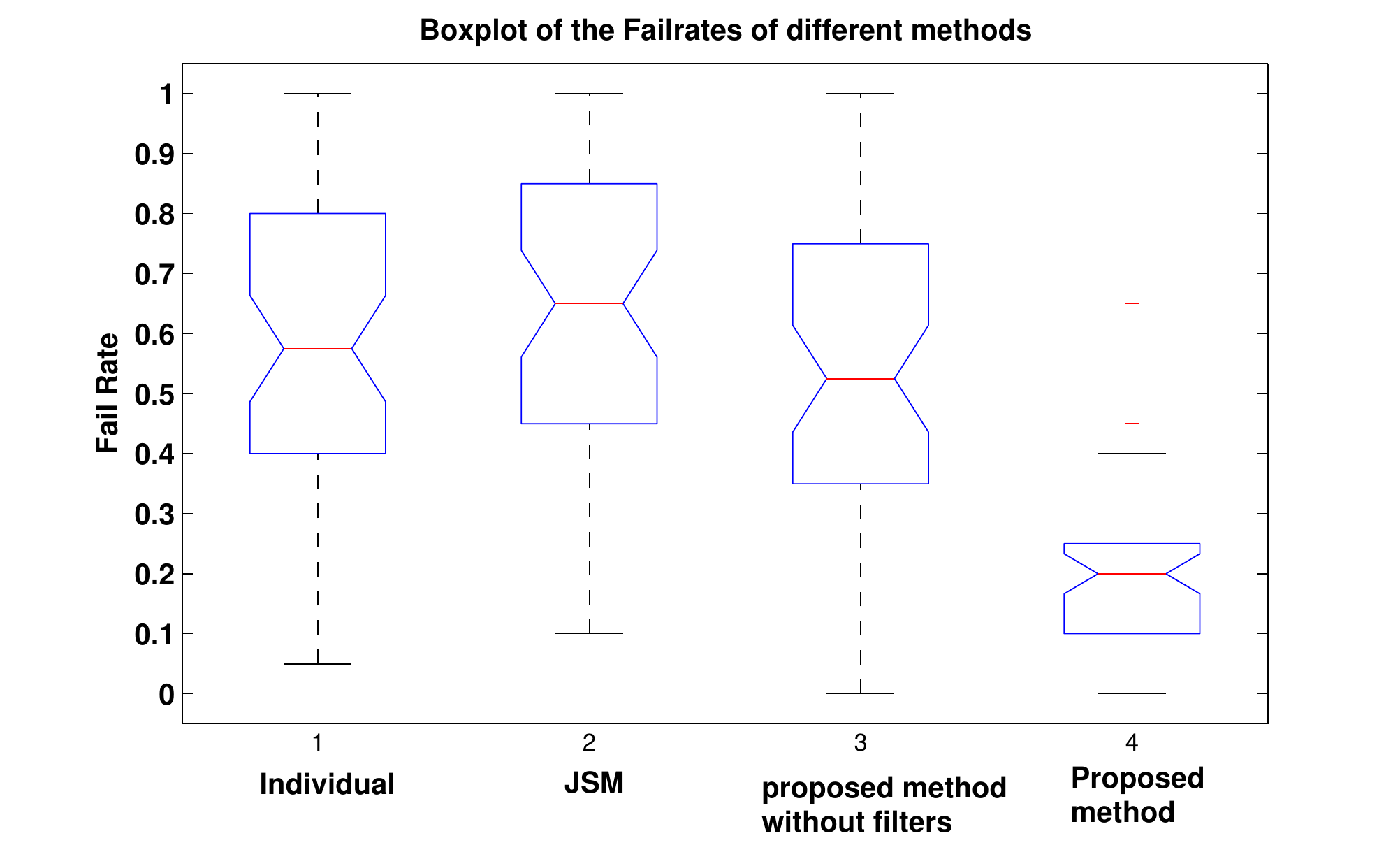}
 \caption{Boxplot of the fail-rate of the reconstruction for the different criteria.}
  \label{boxplot2} 
\end{figure}

\begin{figure}[t]
  \centering
  \includegraphics[width=9cm, height=6cm]{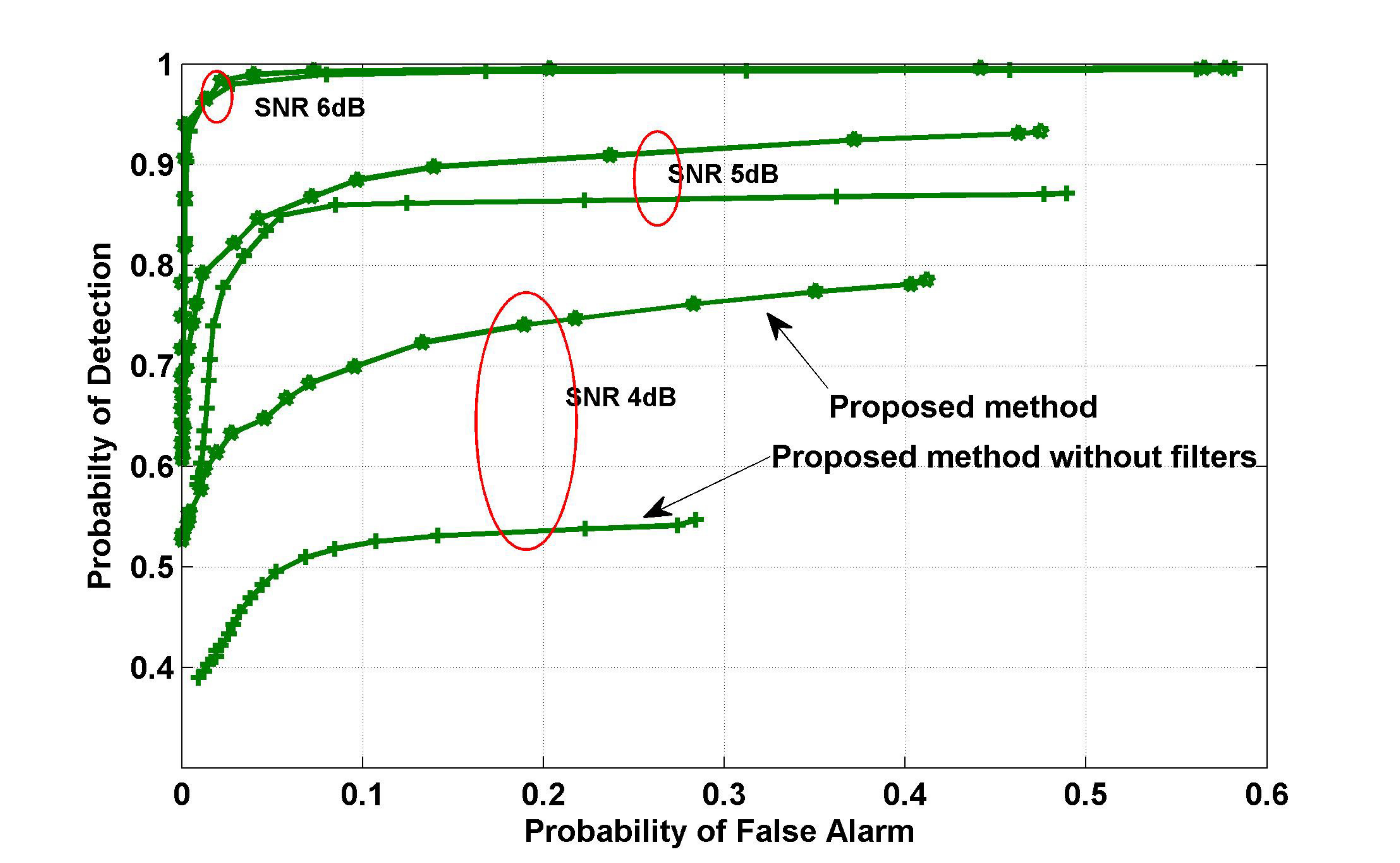}
  \caption{Improvements caused by using and computing disturbance filters in the model for lossy networks.}
  \label{RocBER}
\end{figure}

The impact of using destructive filters in the proposed method can be more inferred by considering Figs. \ref{boxplot2} and \ref{RocBER}.  Figure \ref{boxplot2} shows the boxplot of the fail-rates of the reconstruction equations of the different methods in all of the investigated BERs and the sensing rates.  It can be seen in Fig. \ref{boxplot2} that using the destructive filters make the compressive sensing based reconstruction optimization problem be more feasible and solvable. Alternatively, Fig. \ref{RocBER} compares the ROC curves of the detection performance between the proposed method with and without using the destructive filters for three different SNRs. Figure \ref{RocBER} implies that using the destructive filters in the proposed framework, brings more accuracy of the spectrum sensing especially in higher bit error rates (lower SNRs).

%=======================
\section{Discussion and Conclusions}\label{Conclusion}
%=======================
Since the CR sensors are distributed in the region of support, in order to construct the PSD maps, the sensed PSD by each sensor should be transmitted to a FC. Therefore, when the number of sensors is large, transmitting this type of overhead data can be challenging. In this paper some criteria are proposed based on using shared part of the signals to compress the PSDs and reconstruct them robustly when the data transmissions to the FC is imperfect. Therefore, the perturbation of the transmission system are modeled in the framework by using disturbance filters between each sensor node and the FC. Moreover, the estimation approach for mentioned disturbance filters is proposed. In addition, due to compressing the PSDs of the sensors and therefore the lower transmitted data to send, the proposed method can be used in lower bandwidth usage too. Furthermore, since the compressive sensing sampling method is used to compress the signals of the sensors, using the proposed framework brings lower computational cost and therefore more lifetime in the sensor side too.

\bibliographystyle{IEEEtran}

\begin{thebibliography}{1}

\bibitem{Long}
F.~Bader, P.~Demestricas, L.~DaSilva, and H.~Harada, ``Cognitive radio in emerging communications systemssmall cells, machine-to-machine communications, TV white spaces and green radios," \textit{Transactions on Emerging Telecommunications Technologies}, vol. 24, no. 7-8, pp. 633–635, 2013.

\bibitem{CRSurvey}
T.~Yucek and H.~Arslan, ``A survey of spectrum sensing algorithms for cognitive radio applications," \textit{IEEE communications surveys \& tutorials}, vol. 11, no. 1, pp. 116–130, 2009.

\bibitem{Zhao}
Q.~Zhao and B.~M.~Sadler, ``A survey of dynamic spectrum access," \textit{IEEE signal processing magazine}, vol. 24, no. 3, pp. 79–89, 2007.

\bibitem{GIANNAKIS}
J.~A.~Bazerque and G.~B.~Giannakis, ``Distributed spectrum sensing for cognitive radio networks by exploiting sparsity," \textit{IEEE Transactions on Signal Processing}, vol. 58, no. 3, pp. 1847–1862, 2010.

\bibitem{LASSO}
R.~Tibshirani, ``Regression shrinkage and selection via the lasso," \textit{Journal of the Royal Statistical Society. Series B (Methodological)}, pp. 267–288, 1996.

\bibitem{cooperative}
I.~F.~Akyildiz, B.~F.~Lo, and R.~Balakrishnan, ``Cooperative spectrum sensing in cognitive radio networks: A survey," \textit{Physical communication}, vol. 4, no. 1, pp. 40–62, 2011.

\bibitem{ghasemi2005collaborative}
A.~Ghasemi and E.~S.~Sousa, ``Collaborative spectrum sensing for opportunistic access in fading environments," in \textit{the First IEEE International Symposium on New Frontiers in Dynamic Spectrum Access Networks (DySPAN)}, 2005, pp. 131–136.

\bibitem{CS}
D.~L.~Donoho, ``Compressed sensing," \textit{IEEE Transactions on Information Theory}, vol. 52, no. 4, pp. 1289–1306, 2006.

\bibitem{MyIEICE}
J.~A.~Jahanshahi, M.~Eslami, and S.~A.~Ghorashi, ``PSD map construction scheme based on compressive sensing in cognitive radio networks," \textit{IEICE Transactions on Communications}, vol. 95, no. 4, pp. 1056–1065,
2012.

\bibitem{CrJSM1}
Y.~Wang, A.~Pandharipande, Y.~L.~Polo, and G.~Leus, ``Distributed compressive wide-band spectrum sensing," in \textit{IEEE Information Theory and Applications Workshop}, 2009, pp. 178–183.

\bibitem{CrJSM2}
J.~Liang, Y.~Liu, W.~Zhang, Y.~Xu, X.~Gan, and X.~Wang, ``Joint compressive sensing in wideband cognitive networks," in \textit{IEEE Wireless Communications and Networking Conference (WCNC)}, 2010, pp. 1–5.

\bibitem{Mine}
M.~Eslami, F.~Torkamani-Azar, and E.~Mehrshahi, ``A centralized PSD map construction by distributed compressive sensing," \textit{IEEE Communications Letters}, vol. 19, no. 3, pp. 355–358, 2015.

\bibitem{CrJSM5}
P.~Zhang and R.~Qiu, ``Cooperative wideband spectrum sensing for the centralized cognitive radio network," \textit{arXiv preprint arXiv:1102.3755}, 2011.

\bibitem{CrJSM4}
H.~Sun, ``Collaborative spectrum sensing in cognitive radio networks," \textit{Ph.D. Thesis from The University of Edinburgh}, 2011.

\bibitem{Eslami2017ICASSP}
M.~Eslami, F.~Torkamani-Azar, and E.~Mehrshahi, ``A compressive method for centralized psd map construction," in \textit{IEEE International Conference on Acoustics, Speech and Signal Processing (ICASSP), Accepted for presentation in Ph.D. Forum.}, 2017, https://arxiv.org/abs/1612.02892.

\bibitem{CrCs3}
D.~Sundman, S.~Chatterjee, and M.~Skoglund, ``On the use of compressive sampling for wide-band spectrum sensing," in \textit{IEEE International Symposium on Signal Processing and Information Technology (ISSPIT)}, 2010, pp. 354–359.

\bibitem{BPDN}
P.~R.~Gill, A.~Wang, and A.~Molnar, ``The in-crowd algorithm for fast basis pursuit denoising," \textit{IEEE Transactions on Signal Processing}, vol. 59, no. 10, pp. 4595–4605, 2011.

\bibitem{DCS}
D. Baron, M. B. Wakin, M. F. Duarte, S. Sarvotham, and R. G. Baraniuk, ``Distributed compressed sensing," \textit{IEEE Transactions on Information Theory}, vol. 52, no. 12, pp. 5406-5425, 2006.

\bibitem{BPDN2}
D.~L.~Donoho, et al., ``Sparselab," \textit{SparseLab toolbox shared online, http://sparselab. stanford. edu/, 24th August}, 2007.

\end{thebibliography}

\end{document}